\documentclass[12pt]{iopart}
\usepackage{bm}
\usepackage{wrapfig,epsfig}
\usepackage{iopams}  

\def\ket#1{ $ \left\vert  #1   \right\rangle $ }  
\def\ketm#1{  \left\vert  #1   \right\rangle   }  
  
\def\bram#1{  \left\langle  #1   \right\vert   }  
  
\def\sprm#1#2{  \left\langle #1 \left\vert \right. #2 \right\rangle   }  
 
\def\mem#1#2#3{  \left\langle #1 \left\vert  #2 \right\vert #3 \right\rangle   } 

\def\ninejm#1#2#3#4#5#6#7#8#9{  \left\{ \begin{array}{ccc} 
                                        \        #1 & #2 & #3  \\  
                                                 #4 & #5 & #6  \\
						 #7 & #8 & #9  
				        \end{array} 		 
						 \right\}   }  
\def\twobytwo#1#2#3#4{  \left( \begin{array}{cc} 
                                              #1 & #2 \\[0.2cm]
                                              #3 & #4   
				\end{array} \right)   }

\begin{document}

\title[]{Angular distribution studies on the two-photon
ionization of hydrogen-like ions: Relativistic description\footnote{This
is the preprint of the original: Peter Koval, Stephan Fritzsche and
Andrey Surzhykov, 2004 {\it J. Phys. B: At. Mol. Opt. Phys.} {\bf 37} 375--388.}}

\author{Peter Koval, Stephan Fritzsche and
Andrey Surzhykov\footnote[1]{To whom correspondence should be 
addressed (surz@physik.uni-kassel.de)}}

\address{Fachbereich Physik, Universit\"at Kassel, Heinrich-Plett Str. 40, 
         D-34132 Kassel, Germany}


\begin{abstract}
The angular distribution of the emitted electrons, following the two-photon
ionization of the hydrogen-like ions, is studied within the framework of
second order perturbation theory \textit{and} the Dirac equation. Using a
density matrix approach, we have investigated the effects which arise from 
the polarization of the incoming light as well as from the higher multipoles 
in the expansion of the electron--photon interaction. For medium- and high-Z
ions, in particular, the non-dipole contributions give rise to a significant 
change in the angular distribution of the emitted electrons, if compared 
with the electric-dipole approximation. This includes a strong forward 
emission while, in dipole approxmation, the electron emission always occurs
symmetric with respect to the plane which is perpendicular to the photon
beam. Detailed computations for the dependence of the photoelectron 
angular distributions on the polarization of the incident light are 
carried out for the ionization of H,  Xe$^{53+}$, and U$^{91+}$
(hydrogen-like) ions.
\end{abstract}

\section{Introduction}

During the last decades, the multi-photon ionization of atoms and ions has
been widely studied, both experimentally and theoretically. While, however, the
majority of experiments were first of all concerned with the multi-photon 
ionization of complex atoms, most theoretical investigations instead dealt 
with the ionization (and excitation) of the much simpler hydrogen-like and 
helium-like systems. For atomic hydrogen, in contrast, multi-photon 
experiments have been carried out only recently (Wolff \etal 1988, 
Rottke \etal 1990, Antoine \etal 1996) because of the former lack of 
sufficiently intensive (and coherent) light sources in the UV and EUV 
region. With the recent progress in the setup of intensive light
sources in the EUV and x-ray domain, such as the fourth-generation 
synchrotron facilities or variously proposed free-electron lasers, 
two- and multi-photon studies on the ionization of inner-shell 
electrons are now becoming more likely to be carried out in the future, 
including case studies on medium-Z and high-Z \textit{hydrogen-like} 
ions (Kornberg \etal 2002). With increasing charge (and intensity of 
the light), of course, relativistic effects will become important and have 
been investigated in the past for the two-photon excitation and decay 
(Goldman and Drake 1981, Szymanowski \etal 1997, Santos \etal 2001) 
as well as ionization (Koval \etal 2003) of hydrogen-like ions. So far,
however, all of these studies were focused on the total (excitation or 
decay) rates and ionization cross sections while, to the best of our 
knowledge, no attempts have been made to analyze the effects of relativity on 
\textit{angular} resolved studies.

In this contribution, we explore the angular distribution of the electrons 
following the two-photon ionization of hydrogen-like ions. Second-order 
perturbation theory, based on Dirac's equation, is applied to calculate the 
two-photon amplitudes including the full (relativistic) electron--photon
interaction. The angular distribution of the photo\-electrons are then derived
by means of the density matrix theory which has been found appropriate
for most collision and ionization processes and, in particular, for
angular-dependent studies (Laplanche \etal 1986). Since, however, the basic 
concepts of the density matrix theory has been presented elsewhere at various
places (Blum 1981, Balashov \etal 2000), we will restrict ourselves to rather
a short account of this theory in Subsection 
\ref{density_matrix_general_relations}. Apart from a few basic relations, here
we only show how the angular distribution of the electrons can be traced back 
to the two-photon transition amplitudes. The evaluation of these amplitudes
in second-order perturbation theory and by means of Coulomb--Green's functions
are discussed later in Subsections \ref{transition_amplitude_direct_summation}
and \ref{greens_function}, and including the full decomposition of the photon 
field in terms of its multipole components in Subsection 
\ref{excact_formulation}. 
Using such a decomposition, we have calculated the electron angular 
distributions for the two-photon ionization of the $1s$ ground state of 
hydrogen (H) as well as hydrogen-like xenon (Xe$^{53+}$) and uranium 
(U$^{91+}$). By comparing the angular distributions for different nuclear
charges $Z$, we were able to analyze both, the effects of the polarization 
of the --- incoming --- light and the contributions from higher 
(i.e.\ non-dipole)
multipoles in the decomposition of the electron--photon interaction. These
results are displayed in Section \ref{results} and clearly show that, with
increasing charge $Z$, the higher multipole components lead
to a strong emission in forward direction (i.e.\ parallel to the propagation of
the light), while the electric-dipole approximation alone gives rise to a
symmetric electron emission around the polar angle $\theta \,\approx\, 90^o$, 
similar as obtained by nonrelativistic computations (Zernik 1964, 
Lambropoulus 1972, Arnous \etal 1973). Finally, a brief summary on the 
two-photon ionization of medium and high-Z ions is given in Section 
\ref{summary}.

\section{Theory}

\subsection{Density matrix approach}
\label{density_matrix_general_relations}

Within the density matrix theory, the state of a physical system is described
in terms of so-called \textit{statistical} (or \textit{density}) operators 
(Fano 1957). These operators can be consi\-dered to represent, for instance, 
an ensemble of systems which are --- altogether --- in either a \textit{pure} 
quantum state or in a \textit{mixture} of different states with any degree of 
coherence. Then, the basic idea of the density matrix formalism is to 
\textit{accompany} such an ensemble through the collision process, starting 
from a well defined 'initial' state and by passing through one or, possibly, 
several intermediate states until the 'final' state of the collision process 
is attained.

In the two-photon ionization of hydrogen-like ions, the 'initial' state of 
the (com\-bined) system 'ion \textit{\,plus\,} photons' is given by the bound
electron \ket{n_b j_b \mu_b} \textit{and} the two incoming photons, if we 
assume a zero nuclear spin $ I\,=\,0$. For the sake of simplicity, we also
restrict our treatment to the case that both photons will have 
\textit{equal} momentum: ${\mathbf k}_1 = {\mathbf k}_2 = {\mathbf k}$, while
the spin states of the photons may still differ from each other and are 
characterized
in terms of the \textit{helicity} parameters $\lambda_1$, $\lambda_2 = \pm$ 1 
(i.e. by means of their spin projections onto the direction of propagation 
${\mathbf k}$). Of course, the case of equal photon momenta ${\mathbf k}$
correspond to the most frequent experimental setup of the two-photon
ionization of atoms and ions using, for instance, lasers or synchrotron 
radiation sources. With these assumptions in mind, the initial spin state 
of the overall system is determined by the direct product of the statistical 
operators of the ion and the two incident photons 
\begin{equation}
   \label{initial_density_operator}
   \hat{\rho}_i = 
   \hat{\rho}_b \otimes \hat{\rho}_{\gamma} \otimes \hat{\rho}_{\gamma}
\end{equation}
or, explicitly, in a representation of the density matrix in terms of the
individual momenta by
\begin{eqnarray}
   \label{initial_density_matrix}
   &   & \hspace*{-1.8cm}
   \mem{n_b j_b \mu_b, {\mathbf k} \lambda_1, {\mathbf k} \lambda_2}
   {\hat{\rho}_i}
   {n_b j_b \mu'_b, {\mathbf k} \lambda'_1, {\mathbf k} \lambda'_2} 
   \nonumber \\[0.2cm] \hspace*{1.0cm}
   & = &
   \mem{n_b j_b \mu_b}{\hat{\rho}_b}{n_b j_b \mu'_b} \:
   \mem{{\mathbf k} \lambda_1}{\hat{\rho}_{\gamma}}{{\mathbf k} \lambda'_1} \,
   \mem{{\mathbf k} \lambda_2}{\hat{\rho}_{\gamma}}{{\mathbf k} \lambda'_2} 
   \, .
\end{eqnarray}

In the 'final' state of the ionization, after the electron has \textit{left} 
the nucleus, we just have a free electron with asymptotic momentum 
$\mathbf{p}$ and spin projection $m_s$ (as well as the bare residual ion with
nuclear charge $Z$). Therefore, the final spin state is described by the
statistical operator of the emitted (free) electron $\hat{\rho}_e$ which, in
the framework of the density matrix theory, can be obtained from the 
initial-state density operator $\hat{\rho}_i$  owing to the relation
(Blum 1981, Balashov 2001)
\begin{eqnarray}
   \label{initial_final_operators}
   \hat{\rho}_f & = & \hat{\rho}_e \;=\; 
   \hat{R} \, \hat{\rho}_i \, \hat{R}^+ \, .
\end{eqnarray}
In this simple relation, $\hat{R}$ is called the transition operator and must 
describe the interaction of the electron with the (two photons of the) 
radiation field. Of course, the particular form of the transition operator
$\hat{R}$ depends on the framework in which we describe the coupling of the 
radiation field to the atom. As appropriate for high-Z ions, below we will 
always refer to a \textit{relativistic} treatment of the electron--photon 
interaction, based on Dirac's equation and the \textit{minimal coupling} 
of the radiation field (Berestetskii \etal 1971).

Instead of applying Eq.~(\ref{initial_final_operators}), in practise, it is 
often more convenient to rewrite the statistical operators in a matrix 
representation. Using, for example, the initial spin density matrix
(\ref{initial_density_matrix}), we easily obtain the density matrix of the 
(finally) emitted electron by 
\begin{eqnarray}
   \label{final_density_matrix_general}
   \fl
   \mem{{\mathbf p} m_s}{\,\hat{\rho}_e\,}{{\mathbf p} m'_s} & = & 
   \sum\limits_{\mu_b \mu'_b} \:
   \sum\limits_{\lambda_1 \lambda'_1 \lambda_2 \lambda'_2} \:
   \mem{n_b j_b \mu_b}{\,\hat{\rho}_b\,}{n_b j_b \mu'_b} \,
   \mem{{\mathbf k} \lambda_1}{\,\hat{\rho}_{\gamma}\,}{{\mathbf k} \lambda'_1}
   \, \
   \mem{{\mathbf k} \lambda_2}{\,\hat{\rho}_{\gamma}\,}{{\mathbf k} \lambda'_2}
   \nonumber \\[-0.1cm]
   &   & \hspace*{3.0cm}
   \times \: M_{b {\mathbf p}}(m_s, \mu_b, \lambda_1, \lambda_2) \:
             M^*_{b {\mathbf p}}(m'_s, \mu'_b, \lambda'_1, \lambda'_2),
\end{eqnarray}
where used is made of the abbreviation
\begin{eqnarray}
   \label{transition_matrix_element}
   M_{b {\mathbf p}}(m_s, \mu_b, \lambda_1, \lambda_2) & = &
   \mem{{\mathbf p} \,m_s}{\,\hat{R}\,}{
        {\mathbf k} \lambda_1, {\mathbf k} \lambda_2, n_b j_b \mu_b}
\end{eqnarray}
in order to represent the transition amplitudes for the two-photon
ionization. The final-state density matrix (\ref{final_density_matrix_general})
still contains the \textit{complete} information about the ionization process
(i.e.\ the properties of the bare ion \textit{and} the electron) and, thus, can
be used to derive all the observable properties of the photoelectrons. 
Obviously, however, the outcome of some considered experiment will depend on 
the 
particular setup and the capability of the detectors for \textit{resolving}  
the individual properties of the particles. In the density matrix theory, this  
setup of the experiment is typically described in terms of a (so-called)
\textit{detector operator} $\hat{P}$ which characterizes the detector system
as a whole. In fact, this detector operator can be considered to project out 
all those quantum states of the final-state system which leads to a 'count' 
at the  detectors; in the language of the density matrix, therefore, the 
probability for an 'event' at the detector is simply given by the trace of 
the detector operator with the density matrix: 
$\,W \,=\, {\rm Tr} (\hat{P}\,\hat{\rho})\,$.

%
%
%
%
\begin{center}
\begin{figure}
\hspace*{3cm}
\epsfig{file=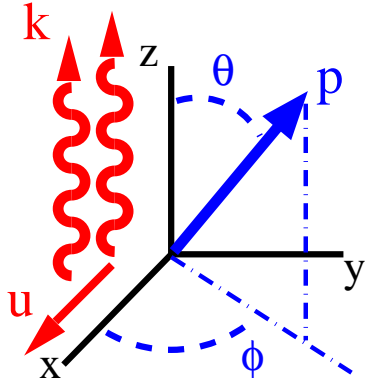, height=5.0cm}
\caption{Geometry of the two-photon ionization. The photoelectron is emitted
along the unit vector $\hat{\mathbf{p}} \:= \: (\theta, \phi)$ where 
$\theta$ is the (polar) angle between the incident photon momenta ${\mathbf k}$
(chosen as the $z$--axis) and the electron momentum ${\mathbf p}$. Moreover,
the (azimuthal) angle $\phi$ defines the angle of ${\mathbf p}$ with respect to
the $x$--$z$ plane which, in the case of linearly-polarized light,
contains the polarization vector ${\mathbf u}$.}
\label{Fig_geometry} 
\end{figure}
\end{center}

To determine, for instance, the angular distribution of the emitted (photo-) 
elec\-trons, we may assume a detector operator in a given direction
$\hat{p} \:=\: (\theta, \phi)$ [cf.\ Figure \ref{Fig_geometry}] which is
\textit{insensitive} to the polarization of the electrons
\begin{equation}
   \label{detector_operator}
   \hat{P} = \sum\limits_{m_s} \ketm{{\mathbf p} m_s} \bram{{\mathbf p} m_s},
\end{equation}
i.e.\ a projection operator along $\mathbf{p}$ and including a summation 
over the spin state $m_s$ of the electrons. From this operator, and by taking 
the trace over the product $(\hat{P}\,\hat{\rho})\,$ with the final-state 
density matrix (\ref{final_density_matrix_general}), we obtain immediately 
the electron angular distribution in the form
\begin{eqnarray}
   \label{angular_distribution_general}
   &  & \hspace*{-2.3cm}
   W(\hat{p}) \; = \; {\rm Tr} (\hat{P}\,\hat{\rho}_f) 
   \; = \; \frac{1}{2j_b + 1} \,
   \sum\limits_{\mu_b m_s} \ 
   \sum\limits_{\lambda_1 \lambda'_1 \lambda_2 \lambda'_2} \:  
   \mem{{\mathbf k} \lambda_1}{\hat{\rho}_{\gamma}}{{\mathbf k} \lambda'_1} \:
   \mem{{\mathbf k} \lambda_2}{\hat{\rho}_{\gamma}}{{\mathbf k} \lambda'_2}
   \nonumber \\[0.1cm]
   &   & \hspace*{4.5cm}
   \times \: M_{b {\mathbf p}}(m_s, \mu_b, \lambda_1, \lambda_2) \:
   M^*_{b {\mathbf p}}(m_s, \mu_b, \lambda'_1, \lambda'_2) \, 
\end{eqnarray}
where, for the sake of simplicity, we have assumed that the hydrogen-like ion 
is initially unpolarized. Apart from this additional assumption, however,
Eq.\  (\ref{angular_distribution_general}) still represents the general form 
of the electron angular distribution for the process of the two-photon 
ionization of hydrogen-like ions.
As seen from this equation, the emission of the photoelectron will depend on the
spin state of the incident photons, defined by the photon density matrices
$\mem{{\mathbf k} \lambda}{\hat{\rho}_{\gamma}}{{\mathbf k} \lambda'}$.
For any further evaluation of this distribution function, therefore, we shall 
first specify  these density matrices or, in other words, the polarization of 
the incoming light. For example, if both photons are \textit{unpolarized}, 
the (two) photon density matrices simply reduce to a constant 1/2,  
$\,\mem{{\mathbf k} \lambda}{\,\hat{\rho}_{\gamma}\,}{{\mathbf k} \lambda'} 
 \:=\: \delta_{\lambda \lambda'}/2\,$ [cf.\ Appendix, Eq.\ 
 \ref{photon_density_matrix}] and leads us to the well-known angular 
 distribution
\begin{eqnarray}
   \label{angular_distribution_unpolarized_photons}
   W^{\rm\, unp}(\hat{p}) & = & 
   \frac{1}{4 (2j_b + 1)} \sum\limits_{\mu_b m_s}
   \: \sum\limits_{\lambda_1 \lambda_2} \,
   \left|\, M_{b \mathbf{p}}(m_s, \mu_b, \lambda_1, \lambda_2) \,\right|^2 .
\end{eqnarray}

For many (modern) light sources such as lasers or synchrotron radiation,
it is not very practical to consider only \textit{unpolarized} light from the
very beginning. In general, instead, the angular distribution of the emitted
electrons will depend both on the \textit{type} as well as the \textit{degree}
of the polarization of the incident light. For \textit{circularly} polarized 
light with degree $P_C$, for instance, the photon density matrix from Eq.\
(\ref{angular_distribution_general})  becomes
$\,\mem{{\mathbf k} \lambda}{\,\hat{\rho}_{\gamma}\,}{
        {\mathbf k} \lambda'} \:=\:
   (1 \,+\, \lambda P_C) \, \delta_{\lambda \lambda'}/2\,$  and, hence, give
rise to the angular distribution
\begin{eqnarray}
   \label{angular_distribution_circular_polarized_photons}
   &  & \hspace*{-2.1cm}
   W^{\rm\, circ}_{P_C} (\hat{p}) \; = \; 
   \frac{1}{4 (2j_b + 1)} \, \sum\limits_{\mu_b m_s \lambda_1 \lambda_2} \,
   (1 \,+\, \lambda_1 \, P_C)  \  (1 \,+\, \lambda_2 \, P_C) \
   \left|\, M_{b \mathbf{p}}(m_s, \mu_b, \lambda_1, \lambda_2) \,\right|^2 \ ,
\end{eqnarray}
while, for \textit{linearly} polarized light along the $x$--axis and with a
polarization degree $P_L$, the photon density matrix is
$\mem{{\mathbf k} \lambda}{\,\hat{\rho}_{\gamma}\,}{{\mathbf k} \lambda'} \:=\:
 \delta_{\lambda \lambda'}/2 \:+\: (1 - \delta_{\lambda \lambda'}) P_L/2$.
If we evaluate Eq.\  (\ref{angular_distribution_general}) again with this
latter density matrix, we obtain the angular distribution 
\begin{eqnarray}
   \label{angular_distribution_linear_polarized_photons_b}
   &  & \hspace*{-2.1cm}
   W^{\rm\, lin}_{P_L}(\hat{p}) \; = \; \frac{1}{4 (2j_b + 1)} \,
   \sum\limits_{\mu_b m_s} \,
   \left( (1 - P_L)^2 \, \sum\limits_{\lambda_1 \lambda_2} \,
   \left| \,M_{b \mathbf{p}}(m_s, \mu_b, \lambda_1, \lambda_2) \,\right|^2 
   \right.
   \nonumber \\[-0.2cm]   
   &  & \hspace*{6.0cm}
   P^2_L \, \left. \left| \, \sum\limits_{\lambda_1 \lambda_2} 
   M_{b \mathbf{p}}(m_s, \mu_b, \lambda_1, \lambda_2)  \, \right|^2 \right)
\end{eqnarray}
for the electrons as emitted in the two-photon ionization of hydrogen-like
ions with linearly polarized light.

\subsection{Two-photon transition amplitude in second-order perturbation
            theory}
\label{transition_amplitude_direct_summation}

For any further analysis of the electron angular distributions, following
the two-photon ionization of a hydrogen-like ion, we need to calculate the 
transition amplitude $M_{b {\mathbf p}}(m_s, \mu_b, \lambda_1, \lambda_2)$ 
as seen from Eqs.\  
(\ref{angular_distribution_unpolarized_photons})--(\ref{angular_distribution_linear_polarized_photons_b}).
This amplitude describes a \textit{bound--free} transition of the electron
under the (simultaneous) absorption of two photons. For a moderate intensity
of the photon field, of course, this amplitude is most simply calculated 
by means of second-order perturbation theory (Laplanche \etal 1976)
\begin{eqnarray}
   \label{transition_amplitude_second_order}
   &  & \hspace*{-2.1cm}
   M_{b \mathbf{p}}(m_s, \mu_b, \lambda_1, \lambda_2) 
   \; = \; \sum\mkern-22mu\int_{\nu} \
   \frac{\mem{{\psi_{{\mathbf p} m_s}}}{{\mathbf \alpha} \,
   { \mathbf u}_{\lambda_1} \, e^{i {\mathbf kr}} }{\psi_\nu}
   \mem{\psi_\nu}{{\mathbf \alpha} \,
   { \mathbf u}_{\lambda_2} \, e^{i {\mathbf kr}} }
   {\psi_{n_b j_b \mu_b}}}{E_{\nu} - E_b - E_{\gamma}} ,
\end{eqnarray}
where the transition operator 
$\mathbf{\alpha} \,\mathbf{u}_{\lambda} \,e^{i {\mathbf kr}}$ describes the
(relativistic) electron--photon inter\-action, the unit vector 
$\,\mathbf{u}_{\lambda}\,$ the polarization of the photons, and where the 
summation runs over the complete one-particle
spectrum. From the energy conservation, moreover, it follows immediately
that the energies of the initial bound state, $E_b$, and the final continuum 
state, $E_f$, are related to each other by $E_f \,=\, E_b \,+\, 2 E_\gamma$, 
owing to the energy of the incoming photons, $E_\gamma \,=\, \hbar c k$. 
Although known for a long time, the
\textit{relativistic} form of the transition amplitude 
(\ref{transition_amplitude_second_order}) has been used only recently in
studying multi-photon ionization processes and, in particular, in order to 
calculate the total ionization cross sections along the hydrogen 
isoelectronic sequence (Koval \etal 2003).
In such a relativistic description of the transition amplitude 
(\ref{transition_amplitude_second_order}), the initial state 
$\psi_{n_b j_b \mu_b}(\mathbf{r}) = \sprm{\mathbf{r}}{n_b j_b \mu_b}$ and
the final state 
$\psi_{{\mathbf p} m_s}(\mathbf{r}) = \sprm{\mathbf{r}}{{\mathbf p} m_s}$
are the (analytically) well-known solutions of the Dirac 
Hamiltonian for a bound and continuum electron, respectively
(Berestetskii \etal 1971).

As seen from Eq.\  (\ref{transition_amplitude_second_order}), the evaluation of
the transition amplitude requires a summation over the \textit{discrete} 
(bound) states as well as an integration over the \textit{continuum} of the
Dirac Hamiltonian, $(\psi_{\nu}, E_\nu)$. In fact, such a 'summation' over
the complete spectrum is difficult to carry out explicitly since, in particular
the integration over the continuum requires the calculation of
\textit{free--free} transitions. In the past, therefore, this summation 
has often  been restricted to some small --- 
\textit{discrete} --- basis, assuming that the contribution from the
continuum is negligible. In practise, however, such a limitation seems 
justified only to estimate the behaviour of the cross sections near the
resonances where the ion is rather likely excited by the first photon into some
 --- real --- intermeditate state of the ion from which it is later ionized by
means of a second photon. In the \textit{non-resonant} region of the
photon energies, in contrast, the integration over the continuum may give
rise to a rather remarkable contribution to the total cross section and, hence,
has to be carried out. Apart from a \textit{direct summation} over the 
continuum states, however, it is often more favorable to apply Green's 
functions, at least if these functions can be generated efficiently. 
For hydrogen-like ions, for example, such Green's functions are known 
analytically, both in the nonrelativistic as well as relativistic theory 
(Swainson and Drake 1991).

\subsection{Green's function approach}
\label{greens_function}

As usual, Green's functions are defined as solutions to some inhomogeneous 
(differential) equation
\begin{equation}
   \label{Green_function_equation}
   \left( E - \hat{H} \right) G_E({\mathbf r}, {\mathbf r'}) \;= \; 
   \delta({\mathbf r} - {\mathbf r'})
\end{equation}
where, in our present investigation, $\hat{H}$ refers the Dirac Hamiltonian 
and $E$ denotes the energy of the atom or ion. For realistic systems, 
of course, such Green's functions are not easy to obtain, even
if only \textit{approximate} solutions are needed. However, a formal solution 
is given by (Morse and Feshbach 1953)
\begin{equation}
   \label{Green_function}
   G_E({\mathbf r}, {\mathbf r'}) \ = \ \sum\mkern-22mu\int_{\nu} \  
   \frac{\ketm{\psi_{\nu}({\mathbf r})} \bram{\psi_{\nu}({\mathbf r'})}}
   {E_\nu - E} \, ,
\end{equation}
including a summation (integration) over the complete spectrum (of $\hat{H}$)
as discussed in the previous section. In the two-photon transition amplitude
(\ref{transition_amplitude_second_order}), therefore, we may simply replace 
this summation by the corresponding Green's function 
\begin{equation}
   \label{transition_amplitude_Green}
   \fl
   M_{b \mathbf{p}}(m_s, \mu_b, \lambda_1, \lambda_2) \ = \
   \mem{\psi_{{\mathbf p} m_s}({\mathbf r})}{{\mathbf \alpha} 
   { \mathbf u}_{\lambda_1} e^{i {\mathbf kr}} 
   G_{E_b + E_\gamma}({\mathbf r}, {\mathbf r'}) \ {\mathbf \alpha} 
   { \mathbf u}_{\lambda_2} 
   e^{i {\mathbf kr'}} }{\psi_{n_b \kappa_b \mu_b}({\mathbf r'})} .
\end{equation}

For hydrogen-like ions, the Coulomb--Green's functions from Eq.\  
(\ref{Green_function_equation}) are known analytically today in 
terms of (various) special functions from mathematical physics and, 
in particular, in terms of the confluent hypergeometric function $ _1F_1(a,b,z)$. 
Here, we will not display these functions explicitly but refer the
reader instead to the literature (Swainson and Drake 1991, 
Koval and Fritzsche 2003). For the further evaluation of the transition
amplitudes (\ref{transition_amplitude_Green}) let us note only that, 
also for the one-particle Dirac Hamiltonian, the Coulomb--Green's function 
can be decomposed into a radial and an angular part
\begin{small}
\begin{eqnarray}
   \label{Green_function_expansion}
   \fl
   G_E({\mathbf r}, {\mathbf r'}) & = &
   \ \frac{1}{r r'} \ \sum\limits_{\kappa m} \ 
   \twobytwo{\mathrm{g}^{\,LL}_{E \kappa}(r, r') \ 
   \Omega_{\kappa m}(\hat{r}) \ \Omega^\dagger_{\kappa m}(\hat{r}') }
   {-i \,\mathrm{g}^{\,LS}_{E \kappa}(r, r') \
   \Omega_{\kappa m}(\hat{r}) \ \Omega^\dagger_{-\kappa m}(\hat{r}')}
   {i \,\mathrm{g}^{\,SL}_{E \kappa}(r, r') \
   \Omega_{-\kappa m}(\hat{r}) \ \Omega^\dagger_{\kappa m}(\hat{r}')}
   {\mathrm{g}^{\,SS}_{E \kappa}(r, r') \
   \Omega_{-\kappa m}(\hat{r}) \ \Omega^\dagger_{-\kappa m}(\hat{r}')} ,
\end{eqnarray}
\end{small}
\newline
where the $\Omega_{\kappa m}(\hat{r})$ denote standard Dirac spinors and where
the radial Green's function is given in terms of four components
$\mathrm{g}_{E \kappa}^{TT'}(r, r')$  with $T = L, S$ refering to the 
\textit{large} and \textit{small} components of the associated (relativistic)
wave functions. The computation of the radial Green's function for 
hydrogen-like ions has been described and implemented previously into the 
\textsc{Greens} library (Koval and Fritzsche 2003); this code has been used 
also for the computation of all transition amplitudes and 
(angle-differential) cross sections as shown and discussed below.

\subsection{Exact relativistic formulation of the two-photon amplitude}
\label{excact_formulation}

Eq.\  (\ref{transition_amplitude_Green}) displays the two-photon transition 
amplitude in terms of the (relativistic) wave and Green's functions of
hydrogen-like ions. For the further evaluation of this amplitude, we  
need to decompose both, the photon as well as the free-electron wave 
functions into partial waves in order to make later use of the techniques of
Racah's algebra. As discussed previously for the capture of electrons 
by bare, high-Z ions  (Surzhykov \etal 2002), we first have to decide about
a proper quantization axis ($z$--axis) for this decomposition, depending ---
of course --- on the particular process under consideration. For the 
photoionization of atoms, the only really \textit{prefared} direction of 
the overall system is given by the photon momenta  
${\mathbf k}_1 = {\mathbf k}_2 = {\mathbf k}$ which we adopt as the 
quantization axis below. Then, the multipole expansion of the radiation field
reads as
\begin{eqnarray}
   \label{photon_wave_expansion}
    {\mathbf u}_{\lambda} e^{i \mathbf{kr}} & = & 
    {\mathbf u}_{\lambda} e^{i kz} \; = \; 
    \sqrt{2\pi} \ \sum_{L=1}^{\infty} \ i^L \ [L]^{1/2} \ 
    \left( \ {\mathbf A}_{L \lambda}^{(m)} + 
    i \lambda {\mathbf A}_{L \lambda}^{(e)} \ \right),
\end{eqnarray}
where  $[L] = (2L + 1)$  and the standard notation $\mathbf{A}_{LM}^{(e,m)}$ 
is used for the electric and magnetic multipole fields, respectively. Each 
of these multipoles can be expressed in terms of the spherical Bessel functions
$j_L(kr)$ and the vector spherical harmonics ${\bf T}^{M}_{L, \Lambda}$ of rank
$L$ as (Rose 1957):
\begin{eqnarray}
   \label{field_expression}
   {\mathbf A}^{(m)}_{L M} = j_L(kr) {\mathbf T}^{M}_{L, L}, \nonumber \\
   {\mathbf A}^{(e)}_{L M} = j_{L-1}(kr) \ \sqrt{\frac{L+1}{2L+1}} 
   {\mathbf T}^{M}_{L, L-1} -  j_{L+1}(kr) \ \sqrt{\frac{L}{2L+1}} 
   {\mathbf T}^{M}_{L, L+1} \, .  
\end{eqnarray}

Using the expressions (\ref{photon_wave_expansion}) and 
(\ref{field_expression}) for the photon field, we can rewrite the
two-photon transition amplitude (\ref{transition_amplitude_Green}) in terms
of its \textit{electric-magnetic} components
\begin{eqnarray}
   \label{transition_amplitude_multipole_expansion}
   \fl
   M_{b \mathbf{p}}(m_s, \mu_b, \lambda_1, \lambda_2) \ = \ 
   2 \pi \sum\limits_{L, L'=1}^{\infty} \ \sum\limits_{\Lambda \Lambda'}
   \ i^{L + L'} \ [L, L']^{1/2} \ \xi^{\lambda_1}_{\Lambda L} \
   \xi^{\lambda_2}_{\Lambda' L'} \nonumber \\[0.2cm]
   \times \mem{\psi_{{\mathbf p} m_s}}{{\mathbf \alpha} \ j_{\Lambda}(kr) \
   {\mathbf T}^{\lambda_1}_{L, \Lambda} \
   G_{E_b + E_\gamma}({\mathbf r}, {\mathbf r'}) \ {\mathbf \alpha} \ 
   j_{\Lambda'}(kr') \ {\mathbf T}^{\lambda_2}_{L', 
   \Lambda'}}{\psi_{n_b \kappa_b \mu_b}} , 
\end{eqnarray}
where the coefficients $\xi^{\lambda}_{L \Lambda}$ are defined as
\begin{eqnarray}
   \label{xi_definition}
   \xi^{\lambda}_{L \Lambda} & = &   
   \left\{ \begin{array}{l l} 
      1                                 & \mbox{if } \Lambda = L \\
      i   \lambda \sqrt{\frac{L+1}{2L+1}} & \mbox{if } \Lambda = L - 1 \\
      -i  \lambda \sqrt{\frac{L}{2L+1}}  & \mbox{if } \Lambda = L + 1
          \end{array}   \right.  \, .
\end{eqnarray}
As seen from the expansion (\ref{transition_amplitude_multipole_expansion}),
we can distinguish between different multipole compo\-nents such as
\textsc{e1e1, e1m1, e1e2}, and others owing to the symmetries of the two vector 
spherical harmonics, i.e. due to the particular combination of the 
summation indices $L,L',\Lambda,\Lambda'$ in this expansion. In the
second line of (\ref{transition_amplitude_multipole_expansion}), however, 
the --- electro-magnetic --- multipole matrix elements still contain the wave 
function $\psi_{\mathbf{p} m_s}(\mathbf{r})$ of the free electron with 
well-defined asymptotic momentum $\mathbf{p}$. In another expansion, 
therefore, we have to decompose it into partial waves to allow for a further 
simplification of the two-photon transition amplitude 
(\ref{transition_amplitude_multipole_expansion}). Again, also the expansion of
the free-electron wave will depend on the choice of the quantization axis 
and requires --- by using a quantization along the photon momentum --- 
that we have to carry out a  rotation of the space part of the electron 
wavefunction from the $z$--direction into the $\mathbf{p}$--direction
\begin{eqnarray}
   \label{continuum_electron_decomposition}
   \psi_{{\mathbf p} m_s}({\mathbf r}) & = & 
   4 \pi \sum\limits_{\kappa_f \mu_f} \
   i^{l_f^L} \ e^{-i \Delta_{\kappa_f}} \, 
   \sprm{l_f^L \mu_f - m_s \ 1/2 m_s}{j_f  \mu_f} 
   \nonumber \\ 
   &   & \hspace*{2.0cm} \times \: 
   Y^*_{l_f^L \ \mu_f - m_s}(\hat{p}) 
   \left( \begin{array}{c}
             \mathrm{g}^L_{E \, \kappa_f}(r)    \;
                \Omega_{\kappa_f \mu_f}(\hat{n}) \\[0.2cm]
             i \: \mathrm{g}^S_{E \, \kappa_f}(r) \; 
                \Omega_{ -\kappa_f \mu_f}(\hat{n})
          \end{array}  \right) ,
\end{eqnarray}
where the summation runs over all partial waves $\kappa_f = \pm 1, \pm 2...$,
i.e. over all possible values of the Dirac angular momentum quantum number 
$\kappa_f = \pm (j_f + 1/2)$ for $l_f^L = j_f \pm 1/2$. In this notation, 
the (nonrelativistic angular) momentum $l_f^L$ represents the parity 
of the partial waves and $\Delta_{\kappa_f}$ is the Coulomb phase shift.
Moreover, as seen from expression (\ref{continuum_electron_decomposition}), 
the partial waves
\begin{eqnarray}
   \label{partial_wave}
   \psi_{E \kappa m_s}(\mathbf{r}) =    \left( \begin{array}{c}
             \mathrm{g}^L_{E \, \kappa_f}(r)    \;\Omega_{\kappa_f \mu_f}(\hat{n}) \\[0.2cm]
             i \: \mathrm{g}^S_{E \, \kappa_f}(r) \; \Omega_{ -\kappa_f \mu_f}(\hat{n})
             \end{array}  \right) 
\end{eqnarray}
separate into a radial and an angular parts, where the two radial functions 
$$
\mathrm{g}^L_{E \, \kappa}(r) \equiv P_{E \, \kappa}(r), \qquad
\mathrm{g}^S_{E \, \kappa}(r) \equiv Q_{E \, \kappa}(r) $$ 
are often called the \textit{large} and \textit{small} components and the 
corresponding angular parts
$\Omega_{\kappa_f \mu_f}(\hat{n}) \equiv \ketm{l_f^L j_f \mu_f} =
\sum_{m_l m_s} \sprm{l_f^L \ m_l \ 1/2 \ m_s}{j_f \ \mu_f} 
Y_{l_f^L m_l}(\hat{n}) \ \chi_{1/2 \ m_s}$ and 
$\Omega_{-\kappa_f \mu_f}(\hat{n}) \equiv \ketm{l_f^S j_f \mu_f} =
\sum_{m_l m_s} \sprm{l_f^S \ m_l \ 1/2 \ m_s}{j_f \ \mu_f} 
Y_{l_f^S m_l}(\hat{n}) \ \chi_{1/2 \ m_s} $ are the standard Dirac spin-angular 
functions.

Using the partial-wave decomposition (\ref{partial_wave}) for the
free-electron wave function and a similar expansion 
(\ref{Green_function_expansion}) 
for the Green's functions, we now can carry out the angular integration
in the transition amplitude (\ref{transition_amplitude_multipole_expansion}) 
analytically
\begin{eqnarray}
   \label{transition_amplitude_final}
   \fl
   M_{b}(m_s, \mu_b, \lambda_1, \lambda_2) &  = & 8 \pi^2 
   \sum\limits_{L \Lambda L' \Lambda'} \sum\limits_{\kappa_f \mu_f}
   \sum\limits_{\kappa m T T'} \ i^{L+L'} \ i^{-l_f^L} \ P^T \ P^{T'} \
   e^{i \Delta_{\kappa_f}} \nonumber \\[0.2cm] 
   &   & \hspace*{1.5cm} \times \: 
   [L, L']^{1/2} \ 
   \xi^{\lambda_1}_{L \Lambda} \  \xi^{\lambda_2}_{L' \Lambda'} \
   \sprm{l_f^L \mu_f - m_s \ 1/2 m_s}{j_f \mu_f} \nonumber \\[0.2cm]
   &   & \hspace*{1.5cm} \times \: 
   \mem{\kappa_f l_f^{\overline{T}} \mu_f}{{\mathbf \sigma} \
   {\mathbf T}^{\lambda_1}_{L \Lambda} }{\kappa l^T m} \
   \mem{\kappa l^{T'} m}{{\mathbf \sigma} \
   {\mathbf T}^{\lambda_2}_{L' \Lambda'} }{\kappa_b l_b^{\overline{T'}} \mu_b}
   \nonumber \\[0.2cm]   
   &    & \hspace*{1.5cm} \times \:  
   U^{\,T T'}_{\,\Lambda \Lambda'}(\kappa_f, \kappa, \kappa_b) \ 
   Y_{l_f^L \ \mu_f - m_s}(\hat{n})
\end{eqnarray}
where, apart from the Clebsch--Gordan coefficient 
$\sprm{l_f^L \mu_f - m_s \ 1/2 m_s}{j_f \mu_f}$ and some constant factors,
the angular part of the amplitude is given in terms of the matrix elements 
of the rank $L$ spherical tensor 
${\mathbf \sigma} {\mathbf T}^{\lambda}_{L \Lambda} \, = \,
[Y_{\Lambda} \otimes \sigma]^M_L$. These matrix elements can be simplified
to (Balashov \etal 2000)
\begin{eqnarray}
   \label{angular_integral}
   \mem{\kappa_b l_b^{T} \mu_b}{{\mathbf \sigma} \
   {\mathbf T}^{M}_{L \Lambda} }{\kappa_a l_a^{T'} \mu_a}   
   & = & \sqrt{\frac{3}{2 \pi}} \ [j_a, L, \Lambda, l_b^{T'}]^{1/2} \
   \sprm{j_a \mu_a \ L M}{j_b \mu_b} \nonumber \\
   & \times & \sprm{l_b^{T} 0, \Lambda 0}{l_a^{T'} 0} \
   \ninejm{l_b^{T}}{1/2}{j_b}{\Lambda}{1}{L}{l_a^{T'}}{1/2}{j_a} \, ,
\end{eqnarray}
by using a proper decomposition in terms of the orbital and spin subspaces. 
The radial part of the transition amplitude
(\ref{transition_amplitude_multipole_expansion}) is contained in
(\ref{transition_amplitude_final}) in the (two-dimensional) integrals 
\begin{equation}
   \label{U_integral}
   \fl
   U^{\,TT'}_{\,\Lambda \Lambda'}(\kappa_f, \kappa, \kappa_b)=
   \int \mathrm{g}_{E_f \kappa_f}^{\overline{T}}(r) \ j_{\Lambda}(kr) \
   \mathrm{g}^{TT'}_{E_b + E_\gamma \kappa}(r,r') \ j_{\Lambda'}(kr') \ 
   \mathrm{g}_{n_b \kappa_b}^{\overline{T'}}(r') \ dr \ dr' ,
\end{equation}
which combines the various (large and small) components of the bound state, 
the Green's function as well as from the free-electron wave. In this notation, 
again, $T = L, \, S$ and a superscript $\overline{T}$ refers to the conjugate 
of $T$, i.e. $\overline{T} = L$ for $T = S$ and \textit{vice versa}. 
In contrast to the angular integrals (\ref{angular_integral}), the radial
integrals (\ref{U_integral}) have to be computed numerically. In the present
work, all the required integrals for the two-photon transition amplitudes
(\ref{transition_amplitude_final}) are calculated by using the 
\textsc{Greens} (Koval and Fritzsche, 2003) and \textsc{Racah} 
(Fritzsche \etal 2001) programs.

\subsection{Electric dipole approximation}

The transition amplitude (\ref{transition_amplitude_final}) still describes 
the full interaction between the electron and photon fields. With the explicit
summation over all the multipoles of the photon field 
(\ref{photon_wave_expansion}), it includes the so-called 
\textit{retardation effects} or \textit{non-dipole} contributions. In practise,
however, the contributions from the higher multipoles decreases very rapidly
with $L$ and may therefore be neglected; in fact, the computation of these
contributions also become rather tedious because of difficulties with a stable 
procedure for the two-dimensional radial integrals (\ref{U_integral}). 
In many cases, therefore, it seems justified to restrict the summation in 
(\ref{transition_amplitude_final}) to just the (dominant) 
\textit{electric dipole} term with $L = 1$ and $\Lambda = L \pm 1$.
This 'dipole approximation' is valid if the photon wave length is
much larger than the size of the atom, i.e.\ $\,k a_0 \ll 1\,$ where 
$a_0$ is the Bohr radius. For the two-photon ionization, this condition is 
well satisfied for most light ions with, say, $Z < 30$ and for photon energies 
below of the one-photon ionization threshold.

From the general form (\ref{transition_amplitude_final}) of the ionization
amplitude, the electric dipole appro\-ximation is obtained by taking
$L = L' = 1$ and $\Lambda, \Lambda' = 0,2$ which --- owing to the dipole 
selection rules --- then also restricts the summation over $\kappa_f$, 
i.e.\ the allowed partial waves for the free electron. For the $K$-shell 
ionization with (completely) \textit{circularly} 
polarized light, for instance, the final-state electron can only escape in the
$d_{3/2}$ or $d_{5/2}$ states. And, as seen from Eq.\  
(\ref{transition_amplitude_final}), the dipole transition amplitude is 
then indeed defined by the (second-rank) spherical harmonic,
$M_{b \mathbf{p}}(m_s, \mu_b, \lambda, \lambda) \,\propto\,
Y_{2, \ \mu_b - m_s + 2 \lambda}(\hat{p}) \,\sim\, sin^2 (\theta) $
 which (together with
eq.~\ref{angular_distribution_circular_polarized_photons}) leads us to the well-known angular distribution 
\begin{equation}
   \label{angular_distribution_circular_electric_dipole}
   W^{\rm \,circ}(\hat{p}) = c_4 \sin^4 \theta
\end{equation}
of the photoelectrons (Lambropoulos 1972, Arnous \etal 1973). As expected from
the axial symmetry of the overall system 'ion \textit{plus} photons', the
angular distribution (\ref{angular_distribution_circular_electric_dipole}) 
only depends on $\theta$ but not on the azimuthal angle $\phi$. For linearly
polarized light, in contrast, a \textit{reaction plane} is naturally defined by
the photon momentum ${\mathbf k}$ and the pola\-rization vector ${\mathbf u}$ and,
hence, the axial symmetry is broken. For a linear polarization of the incident
light, therefore, the angular distribution will depend on both, the polar 
and azimuthal angle and is given by (Zernik 1964, Lambropoulos 1972)
\begin{equation}
   \label{angular_distribution_linear_electric_dipole}
   W^{\rm \,lin}(\hat{p}) = b_0 + b_2 \sin^2\theta \cos^2\phi + 
                      b_4 \sin^4\theta \cos^4\phi,
\end{equation}
where the angle $\phi$ = 0 corresponds to an electron emission within the 
reaction plane.

\section{Results and discussion}
\label{results}

%
%
%
%
\begin{figure}[t]
\begin{center}
\epsfig{file=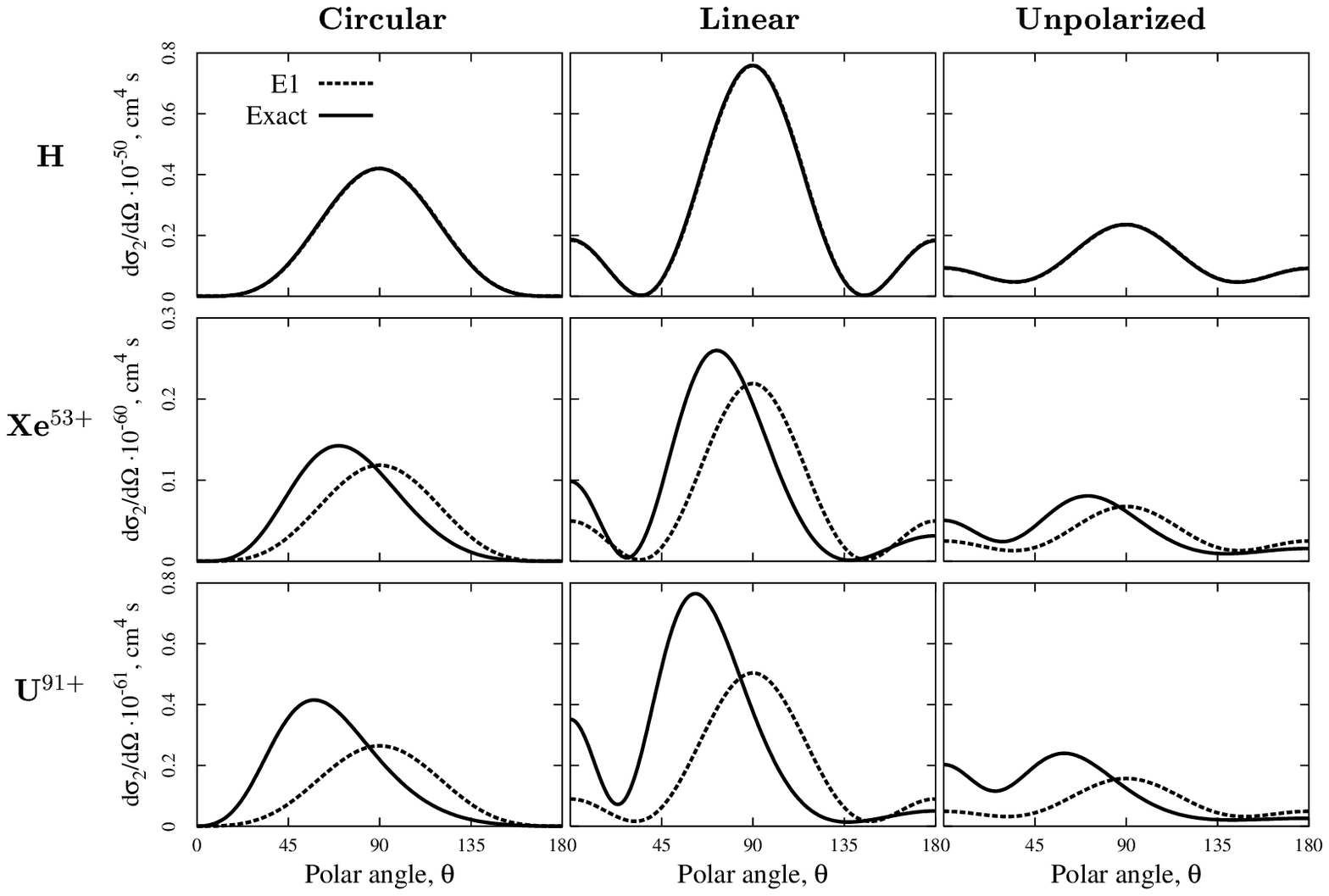, width=15cm, angle=0, clip=
bbllx=10cm, bblly=15cm, bburx=19cm, bbury=26.5cm}
\end{center}
\caption{Angular distributions of the emitted electrons in the two-photon
$K$-shell ionization of hydrogen-like ions by means of circularly, linearly 
and unpolarized light. Results are presented for both, the electric dipole 
(-- --) and the relativistic (---) approximations and for a two-photon 
energy which is 40 \%{} above the (one-photon) ionization threshold.}
\end{figure}

For the calculation of total two-photon ionization cross sections, 
the electric dipole approximation was recently found sufficient for most of the
hydrogen-like ions, and not just in the low-Z domain (Koval \etal 2003). 
Even for high-Z ions, for example, the total cross sections from the dipole 
approximation do not differ more than about 20~\%{} from those of a full 
relativistic computation, including the contributions from all the higher 
multipoles. 
Larger deviations, however, can be expected for the angular distribution
of the emitted electrons which is known to be sensitive to the retardation 
in the electron--photon interaction. As known, for instance, from the 
radiative recombination of high-Z ions, a significant change in the 
angle-differential cross sections may arise from the higher multipoles and 
may lead to quite sizeable deviations when compared with the dipole 
approximation (Surzhykov \etal 2002).

In this contribution, therefore, we have analyzed both, the electric dipole and 
the exact relativistic treatment from Eq.\  (\ref{transition_amplitude_final})
in order to explore the relativistic and retardation effects on the angular
distributions of the electrons. Detailed computations have been carried out, in
particular, for the $K$-shell ionization of (neutral) hydrogen as well as 
hydrogen-like xenon and uranium. Moreover, to explore the dependence
of the relativistic effects on the polarization of the incoming light, three
cases of the polarization are considered: (i) completely circular polarized, 
(ii) completely linear polarized as well as the case
of (iii) unpolarized light. For these three ions and types of polarization, 
Figure 2 displays the angular distributions of the electrons as obtained by the
dipole approximation (-- --) and the exact relativistic treatment (---). While,
for hydrogen, both approximation virtually yields identical results, the start
to differ as the nuclear charge $Z$ is increased. Instead of a symmetrical
emission with respect to the polar angle $\theta$ = 90$^\circ$, then the 
emission
occurs predominantly into forward direction, an effect which is best seen for
hydrogen-like $U^{91+}$ ions. We therefore find, that the \textit{non-dipole}
terms give first of all rise to an \textit{asymmetrical} shift in the
angular distribution of the electrons which could be observed in experiment.
The maxima in the (angle-differential) cross sections, on the other hand,
are less affected and deviate, even for hydrogen-like uranium, less than a
factor of 2.

%
%
%
%
\begin{figure}[t]
\begin{center}
\hspace*{-1cm}
\epsfig{file=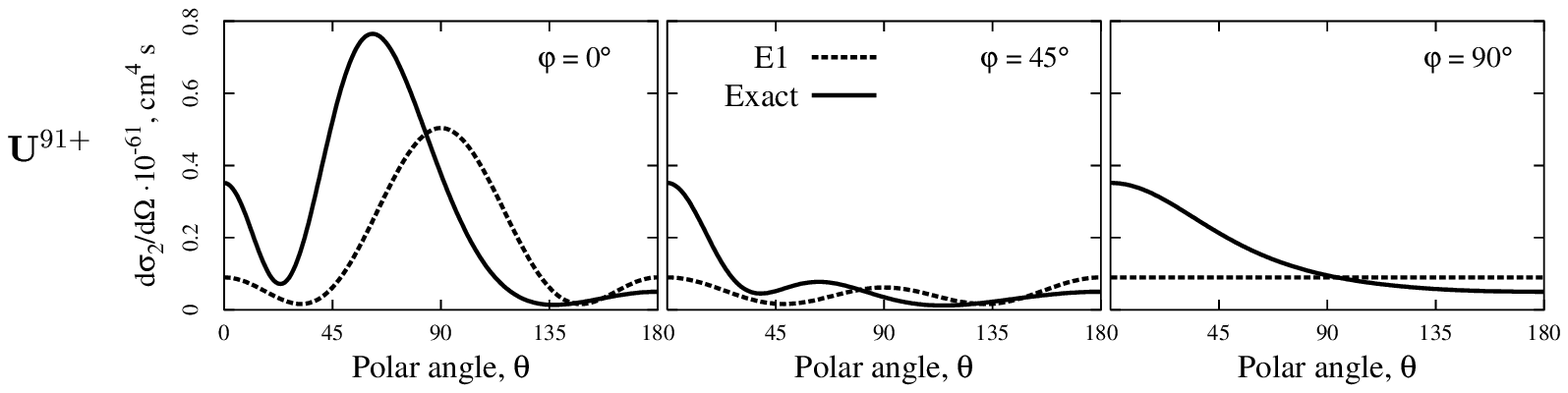, width=16cm, angle=0, clip=
bbllx=10cm, bblly=22cm, bburx=19cm, bbury=26cm}
\end{center}
\caption{Angular distributions of the electrons emitted in the two-photon
$K$-shell ionization of the hydrogen-like uranium U$^{\,91+}$ by means of
linear polarized light. Distributions are shown for the angles 
$\phi$ = 0$^\circ$, 45$^\circ$ and 90$^\circ$ with respect to the reaction
plane; cf.\ Figure 1.}
\end{figure}

In Figure 2, all angular distributions are shown as function of the polar
angle~$\theta$, i.e.\ with respect to the incoming photon beam. As discussed
above, this dependence of the differential cross sections,
$\,\frac{d \sigma}{d \Omega} \,=\, \frac{d \sigma}{d \Omega} (\theta)\,$,
can be the only one for circular and unpolarized light 
for which the electron emission must be axially symmetric. For linear polarized
light, in contrast, the emission of the electrons will depend on both, the 
polar angle~$\theta$ and the azimuthal angle~$\phi$. For this polarization, 
Figure 2 only displays the angular distributions \textit{within} the reaction 
plane, i.e.\  at $\phi$ = 0$^\circ$. To explore, in addition, also the
$\phi$-dependence of the two-photon ionization by linear polarized light
explicitly, Figure 3 shows the corresponding angular distributions 
$\,\frac{d \sigma}{d \Omega} (\theta, \phi)$ for the three particular angles 
$\phi$ = 0$^\circ$, 45$^\circ$ and 90$^\circ$ with respect to the reaction 
plane; here, the left inlet ($\phi$ = 0$^\circ$) is the same as shown 
in Figure~2 in the middle column for U$^{\,91+}$ ions. Again, the results from
the electric dipole approximation are compared with those from a fully 
relativistic computation. As seen from Figure 3, the most pronounced effect 
of the higher multipoles arise for an electron emission in a plane, which is
perpendicular to the photon polarization vector ($\phi$ = 90$^\circ$). 
In such a --- perpendicular --- geometry of the experiment, the cross sections
from the exact treatment show strong forward emission of the photoelectrons 
while the electric dipole approximation
(\ref{angular_distribution_linear_electric_dipole}), in contrast, results in 
a completely isotropic emission, if seen as function of the polar 
angle~$\theta$.

%
%
%
%
\begin{figure}
\hspace*{3cm}
\epsfig{file=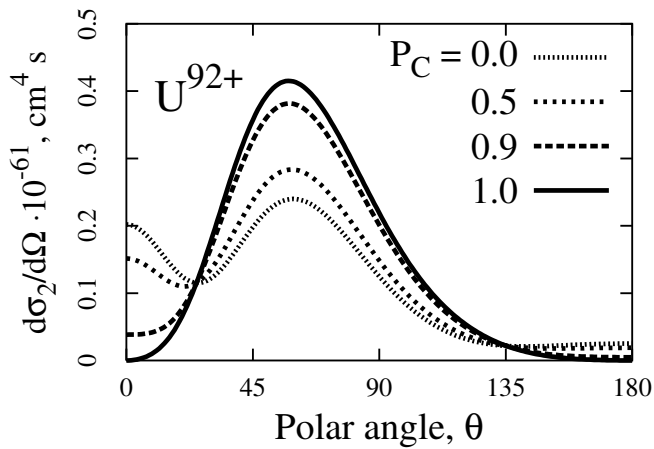, 
        height=9.0cm, angle=270, clip=}
\caption{Angular distributions of the electrons emitted in the two-photon
$K$-shell ionization of the hydrogen-like uranium U$^{91+}$ by circular
polarized light with different degrees of polarizations 
$P_C$ = 0, 0.5, 0.7. and 1.}
\label{Fig_4} 
\end{figure}
%
%

Until now, we considered the two-photon ionization of hydrogen-like ions
by either \textit{completely} polarized (linear: $P_L$ = 1; circular $P_C$ = 1) 
or unpolarized light ($P_L$ = $P_C$ = 0). In most experimental investigations
on two- (and multi-) photon processes, however, the incident radiation is
typically polarized with some given \textit{degree of polarization} 
$0 \,\le\, P_C, \,P_L \,\le\, 1$. Apart from the \textit{type} of the
polarization of the in\-coming light, therefore, we shall study also how the
angular distributions depend on the \textit{degree} or polarization.
Figure 4, for instance, displays the angular distribution from the $K-$shell
ionization of hydrogen-like U$^{\,91+}$ ions by means of circular polarized
light with a degree of polarization $P_C$ = 0.0 (unpolarized case), 0.5, 0.9 
and 1.0. As seen from this figure, the probability for an electron emission
increases at angles around $\theta$ = 60$^\circ$ but decreases (towards zero) 
in forward and backward direction as the degree of polarization is increased.
In particular the behaviour near $\theta$ = 0$^\circ$ and 
$\theta$ = 180$^\circ$ can be easily explained if we consider the conservation
of momentum in the overall system. Since, for completely circular polarized 
light the (total) angular momentum of photons on the quantization axis 
(which is chosen along the photon momenta $\bm{k}$)
becomes
$\lambda_1 + \lambda_2 = \pm $ 2, it obviously can not be compensated ---
in the final state --- if the electron is emitted parallel 
(or antiparallel) to the incoming light,  and hence its spin projection 
is $\mu_f=m_s=\pm 1/2$.
For \textit{unpolarized} light, in contrast, the photons may have different 
helicities and, hence, their angular momentum $\lambda_1 + \lambda_2 = \pm $ 0 
will be conserved under a forward and backward \textit{non spin-flip} 
electron emission.

\section{Summary}
\label{summary}

In this paper, the two-photon ionization of hydrogen-like ions has been
studied in the framework of second-order perturbation theory \textit{and}
the relativistic description of the electron and photon fields. That is,
exact Dirac bound and continuum wave functions were applied for the 
description of the electron to reveal the importance of \textit{relativity}
on the angular distributions of the emitted electrons. Moreover,
relativistic Coulomb--Green's functions are used to perform the summation over
the complete Dirac spectrum as needed in second-order perturbation theory.

To understand the angular distributions of the emitted photoelectron and, in
particular, the influence of the polarization of the light on this emission,
density matrix theory has been utilized to 'combine' the two-photon 
transition amplitudes in
a proper way. Calculations are carried out for the $K$-shell ionization of 
the three (hydrogen-like) ions H, Xe$^{\,53+}$ and U$^{\,91+}$. From the
angular distribution of the electrons for different types (linear, circular,
unpolarized) and degrees of polarization (i.e.\ in going from the completely 
polarized to unpolarized light), it is clearly seen that the angular 
emission depends much more sensitive on the contributions from higher multipoles
than the total cross sections. Two rather pronounced effects, for example, 
concern the (asymmetrical) forward emission of the electrons as well as a 
significant change in the electron emission for linear polarized light, if the
electrons are observed perpendicular to the reaction plane [cf.\ Figure 4]. 
Both effects are enhanced if the nuclear charge of the ions is increased.

An even stronger influence from the non-dipole terms (of the radiation field)
is expected for the spin-polarization of the photoelectrons. Similar as in 
the present investigation, density matrix theory provides a very suitable tool 
for such \textit{polarization} studies.  A detailed analysis of the 
polarization of the photoelectrons, emitted in the two-photon ionization 
of hydrogen-like ions, is currently under work.

\appendix
\section{Photon spin density matrix}

A \textit{pure} (i.e.\ completely polarized) state of the photon can
be characterized in terms of a polarization unit vector ${\mathbf u}$ 
which always points perpendicular to the (asymptotic) photon momentum 
${\mathbf k}$. Of course, this polarization vector, ${\mathbf u}$, 
can be rewritten by means of any \textit{two} (linear independent) basis 
vectors such as the \textit{circular polarization} vectors ${\bf u}_{\pm 1}$ 
which are (also) perpendicular to the wave vector ${\mathbf k}$  and 
which, for ${\bf u}_{+1}$ respective ${\bf u}_{-1}$, are associated with
right- and left-circular polarized photons (Blum81). In such a basis, 
the unit vector for the \textit{linear} polarization of the light can be 
written as
\begin{equation}
   \label{u_linear_definition}
   {\mathbf u}(\chi) = \frac{1}{\sqrt{2}} 
   \left( \ e^{-i \chi} \ {\mathbf u}_{+1} +
   e^{i \chi} \ {\mathbf u}_{-1} \right), 
\end{equation}
where $\chi$ is the angle between ${\mathbf u}(\chi)$ and the $x$--$z$ plane.

While a description of the polarization of the light in terms of either the
circular polarization vectors ${\bf u}_{\pm 1}$ or the linear polarization
vector (\ref{u_linear_definition}) is appropriate for completely polarized 
light, it is not sufficient to deal with an ensemble of photons which have
different polarization. Such a --- mixed --- state of the light is then better 
described in terms of the spin--density matrix. Since the photon 
(with spin $S=1$) has only two allowed spin (or helicity) states 
$\ketm{\mathbf{k} \lambda},\; \lambda \,=\,  \pm 1$,
the spin--density matrix of the photon is a $2\times2$ matrix and, hence, 
can be parameterized by three (real) parameters: 
\begin{eqnarray}
   \label{photon_density_matrix}
   \mem{\mathbf{k} \lambda \, }{ \, \hat{\rho}_{\gamma} \, }{  
   \mathbf{k} \lambda^{'} \, } =    
   \frac{1}{2} \, \left( \begin{array}{cc}  
                     1 \,+\, P_{C}        &  P_{L} \ e^{-2 i \chi} \\[0.3cm]  
                     P_{L} \ e^{2 i \chi}   &  1 \,-\, P_{C}  
                  \end{array} \right) \, , 
\end{eqnarray}
where $0 \le P_L \le 1$ and $-1 \le P_C \le 1$ denote the degree of linear and
circular polarization, respectively. The angle~$\chi$, moreover, represents
the direction of the maximal linear polarization of the light.

Of course, the choice of the parameters $P_L,\, P_c$ and $\chi$ is 
not \textit{unique} and many other --- equivalent --- sets of three real 
parameters could be applied to characterize the photon spin density matrix 
(\ref{photon_density_matrix}). In the analysis of experimental data, for
instance, one often uses the three \textit{Stokes} parameters to describe the
polarization of radiation. The Stokes parameters can easily be expressed in
terms of the (two) degrees of polarization, $P_L$ and $P_C$, and the angle 
$\chi$ as:
\begin{eqnarray}
   \label{Stokes_parameters_definition}
   P_1 &  = &  P_L \ \cos \, 2 \chi , \qquad
   P_2 \; = \; P_L \ \sin \, 2 \chi , \qquad
   P_3 \; = \; P_C \, .
\end{eqnarray}
The use of the Stokes parameters leads to the familiar form of the
spin density matrix (Balashov \etal 2000) 
\begin{eqnarray}
   \label{photon_density_matrix_Stokes}
   \mem{\mathbf{k} \lambda \, }{ \, \hat{\rho}_{\gamma} \, }{  
   \mathbf{k} \lambda^{'} \, } =    
   \frac{1}{2} \, \left( \begin{array}{cc}  
                     1 \,+\, P_{3}        &  P_1 - i P_2 \\[0.3cm]  
                     P_1 + i P_2  &  1 \,-\, P_3  
                  \end{array} \right) \, .   
\end{eqnarray}

%
%

\section*{References}
\begin{harvard}
\item[] Antoine P, Essarroukh N--E, Jureta J, Urbain X and Brouillard F 
        1996 \JPB \textbf{29} 5367
\item[] Arnous E, Klarsfeld S and Wane S, 1973 {\it Phys. Rev.} A {\bf 7} 1559
\item[] Balashov V V, Grum--Grzhimailo A N and Kabachnik N M, 2000 {\it 
        Polarization and Correlation Pheno\-mena in Atomic Collisions} 
        (New York: Kluwer Academic Plenum Publishers)
\item[] Berestetskii V B, Lifshitz E M and Pitaevskii L P, 1971 {\it
 Relativistic Quantum Theory} (Oxford: Pergamon)
\item[] Blum K, 1981 {\it Density Matrix Theory and Applications} (New York: 
        Plenum)
\item[] Eichler J and Meyerhof W, 1995 {\it Relativistic Atomic Collisions} 
        (San Diego: Academic Press) 
\item[] Eichler J, Ichihara A and Shirai T, 1998 {\it Phys. Rev.} A {\bf 58}
        2128 
\item[] Fano U and Racah G, 1959 {\it Irreducible Tensorial Sets} (New York:
        Academic Press)
\item[] Fritzsche S, Inghoff T, Bastug T and Tomaselli M, 2001 
        {\it Comput. Phys. Commun.} {\bf 139} 314       
\item[] Goldman S P and Drake G W, 1981 {\it Phys. Rev.} A \textbf{24} 183
\item[] Kornberg M~A, Godunov A~L, Ortiz S~I, Ederer D~L, McGuire J~H and
        Young L 2002 \textit{Journal of Syn\-chrotron Radiation} \textbf{9} 
        298
\item[] Koval P and Fritzsche S, 2003 {\it Comp. Phys. Comm.} 
\item[] Koval P, Fritzsche S and Surzhykov A, 2003 
        {\it J. Phys. B: At. Mol. Opt. Phys.} {\bf 36} 873
\item[] Lambropoulos P, 1972 {\it Phys. Rev. Lett.} {\bf 28} 585 
\item[] Laplanche G, Durrieu A, Flank Y, Jaouen M and Rachman A, 1976
        {\it J. Phys. B: At. Mol. Opt. Phys.} {\bf 9} 1263
\item[] Laplanche G, Jaouen M and Rachman A, 1986 \JPB \textbf{19} 79
\item[] Morse P and Feshbach H, 1953 {\it Methods of Theoretical Physics}, 
        Vol. 1 (McGraw--Hill Inc., New York)
\item[] Rose M E, 1957 \textit{Elementary Theory of Angular Momentum} 
        (Wiley, New York)
\item[] Rottke H, Wolff B, Brickwedde M, Feldmann D and Welge K H, 1990
        Phys. Rev. Lett. {\bf 64} 404
\item[] Santos J P, Patte P, Parente F and Indelicato P, 2001 
        \EJP D \textbf{13} 27
\item[] Surzhykov A, Fritzsche S and St\"ohlker Th, 2002
        {\it J. Phys. B: At. Mol. Opt. Phys.} {\bf 35} 3713
\item[] Swainson R A and Drake G W F, 1991 {\it J. Phys. A: Math. Phys.} 
        {\bf 24} 95
\item[] Szymanowski C, V\'e{}niard V, Ta\"{\i}{}eb R and Maquet A, 1997
        Europhys.~Lett. \textbf{6} 391
\item[] Wolff B, Rottke H, Feldmann and Welge K H, 
        1988 {\it Z. Phys.} D {\bf 10} 35
\item[] Zernik W, 1964 {\it Phys. Rev.} {\bf 135} A51
\end{harvard}

\end{document}